# Multiphysics Analysis of the 2.5 MeV BNCT RFQ Accelerator

Zhu Xiao-Wen(朱晓文)　　Lu Yuan-Rong（陆元荣）Zhu Kun（朱昆）

State Key Lab of Nuclear Physics and Technology, Peking University, Beijing 100847, China

**Abstract:**　Boron Neutron Capture Therapy (BNCT), is an advanced cancer therapy that destroys the cancer tumors using the well-known $^7_3Li(p,n)^7_3Be$. Because of the highly selectively reaction between a boron and a neutron, BNCT is effective for rapidly spreading cancer, invasive carcinoma, such as head and neck cancer, melanoma, malignant brain tumors and so on. The PKU RFQ group proposes an RFQ based neutron source for BNCT application. The 162.5 MHz four-vane RFQ will accelerate 20-mA H+ from 35.0 keV to 2.50 MeV in CW mode, and delivers a neutron yield of $1.73 \times 10^{13} n/sec/cm^2$. The thermal management will become the most important issues. The detailed multiphysics analysis of the BNCT RFQ will be studied, and the RFQ frequency shift during nominal operating condition is also predicted. The multiphysics analysis is performed by using the CST Multiphysics Model and verified with ANSYS Multiphysics.

**Key words**：Multiphysics Analysis, CW BNCT RFQ

## 1 Introduction

Varieties of accelerator-based neutron source for BNCT[1-3] are being developed and constructed worldwide. AB-BNCT[4-5], as a promising alternative, provides great potential for compatible installation and clinical application in a hospital, with simplified licensing and regulations in maintaining the neutron source than a dedicated fission reactor. The 2.5 MeV RFQ accelerates 20-mA proton beams up to 2.5 MeV in 5.2 m. A modest inter-vane voltage (65 kV) is adopted and the Kilpatrick factor is only 1.24, which is rather reliable and easing the thermal management for CW operation. The BNCT RFQ keeps the focusing strength B constant, which merits in fabricating RFQ cavities. Such design makes the cross-section of RFQ constant, and makes it be easier to tune the cavity, and NC shaping miller can be used for machining the RFQ vanes.

The 5.2 m long 4-vane RFQ is divided into 5 models separately. Each of the RFQ modules will be approximately 1.04 m, and will consist of 4 solid vanes, fabricated from OFHC copper. Due to high power, CW operation, brazed copper structure is preferable. 20 pairs of water-cooled pi-rods are incorporated to provide the field stabilization along the RFQ, and a set of 100 fixed tuners is installed for frequency adjustment and local field correction. To mitigate the RF heat load on the inner surface of the RFQ cavity and tune the operating frequency, 12 cooling channels are embedded into the vanes. Detailed multiphysics analysis of the RFQ resonant structure, including cross-section, tuner period, pi-mode stabilizer period, and cut-backs are performed.

## 2 Analysis Procedure

The multiphysics analysis is performed by using the CST Multiphysics Model[6], and verified with ANSYS Multiphysics[7]. The RF thermal loss is firstly calculated in CST MWS and applied to the inner cavity walls of the RFQ body. The temperature map obtained from the CST MP thermal stationary analysis, is piped into the next structural analysis. Eventually, the deformed meshes of the cavity in CST MP structural analysis are used for mesh morphing and imported into the CST MWS for RF sensitivity analysis, which takes into the prediction of the frequency during the nominal operation, and the frequency sensitivity induced from the other parameters, such as the position of the cooling channel, the diameters of the cooling channel, water temperature, velocity and so on. To calculate the power loss of each element in RFQ structure, the electric field strength in the RFQ periods has been always scaled to the nominal inter-vane voltage (65 kV).

The heat transfer coefficient $h$, between the cooling water and the inner surface of the circular pipe can be calculated as following[8];

$$h = \frac{N_u K}{d} \quad (1)$$

where $K$ denotes the thermal conductivity of cooling water, $d$ is the diameter of the cooling channel, and $N_u$ is the Nusselt number[9],

$$N_u = 0.023 R_e^{0.8} P_r^{0.4} \quad (2)$$

The $R_e$ represents the Reynolds number,

$$R_e = \frac{\rho \bar{v} d}{\mu} \quad (3)$$

Here ρ and μ corresponds to the water density and viscosity respectively, $\bar{v}$ is the average velocity of cooling water.

And $P_r$ is the Prandtl number,

$$P_r = \frac{C_p \mu}{K} \quad (4)$$

where $C_p$ is the specific heat of water at constant pressure.

In thermal analysis, the background and boundary conditions are needed to be adjusted. The ambient temperature and the cooling water temperature are both set to be 293.1 K.

## 3 Layout of cooling channel

A 200 mm length RFQ model is used in the simulations. The RF heat load is calculated in CST MWS, which is shown in Fig. 1. 12 Cooling channels with a diameter of 12 mm are embedded into the vanes, walls and corners of RFQ cavity in each module. Table 1 summarizes the parameters of the cooling channels. The layout of the cooling channel design is mainly focused on the optimization of the temperature distribution along the transverse profile of cross-section, which results in the frequency shift generated by the thermal deformation. The temperature along the RFQ transverse profile is shown in Fig. 2, by varying the geometry parameter of vane channel (*Hv*) in Fig. 3, while keeping the parameters of wall channel (*Hw*) and corner channel (*Hc*) constant. The maximum temperature is located at the vane-tips. The minimum temperature is 297.3 K, and is found in the corner cooling passage. When *Hv* equals to 110 mm, the RFQ body has a minimum temperature rises 3.3 - 6.2 K above the nominal water temperature. Similarly, the positions of the wall channel and corner channel can be optimized. So does the diameters of the cooling passage.

Table 1. Summary of the cooling-water channels.

|  | Vane, wall, Corner Channel |
|---|---|
| Diameter (mm) | 12 |
| Flow speed (m/s) | 2.29 |
| Flow rate (l/s) | 0.26 |
| Number of Channels (/unit) | 4/4/4 |
| Heat transfer coefficient (W/m².K) | 9137 |

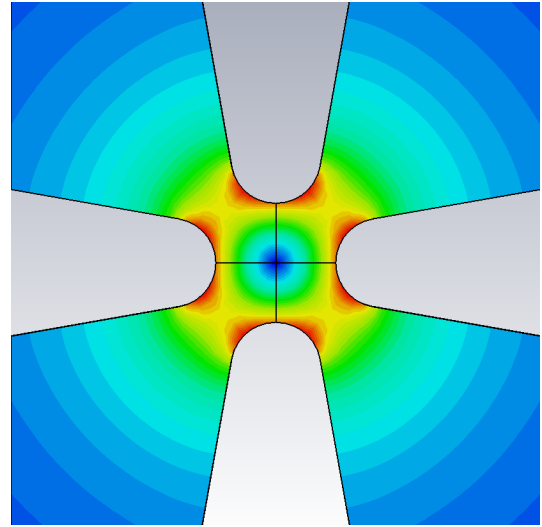

Fig. 1. Electric field near the vane tips.

Fig. 3 gives the temperature contours of the optimal result. Then it is piped into the structure analysis to determine the displacement, which was presented in Fig. 4. The tuning sensitivity of vane temperature is -13.32 kHz/K, and that in wall channel is 10.4 kHz/K.

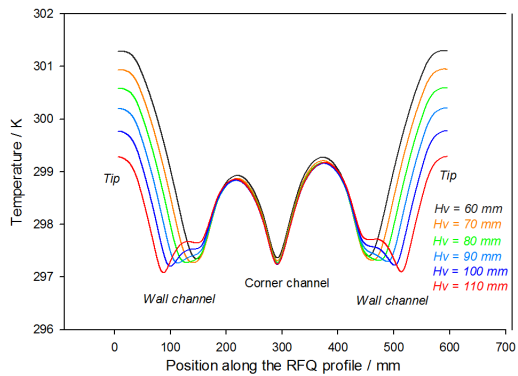

Fig. 2. Temperature distribution along the transverse profile of a quadrant.

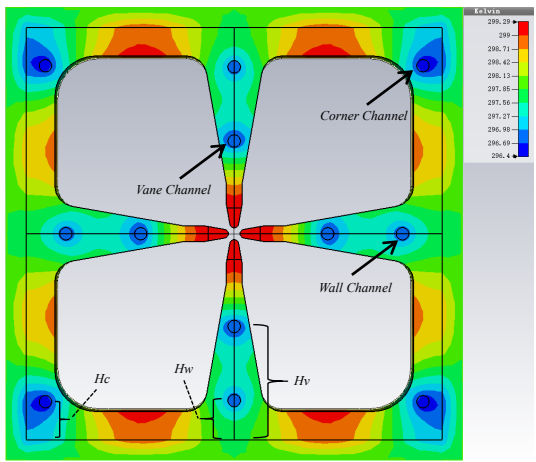

Fig. 3. Temperature contours (K) for steady-state RFQ operation.

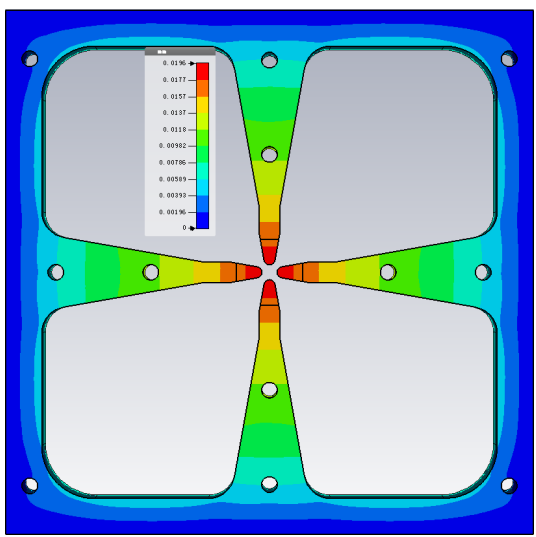

Fig. 4. Thermal displacement due to heat load.

## 4 Cut-back

Once the layout of the cooling channel in the cross-section is determined, the RFQ cut-backs are simulated in a similar way. A short 270 mm cut-back was put into RF analysis. The thermal losses is piped into the thermal analysis, the result of which is presented in Fig. 5. As shown in Fig. 5, the maximum temperature located at the tips of the cut-backs, which is farthest from the cooling passage.

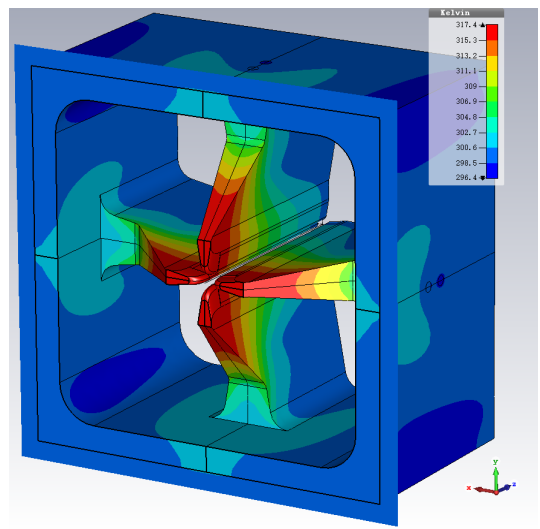

Fig. 5. Cut-back temperature at RFQ entrance.

## 5 Pi-rods

The longitudinal field stabilization in the BNCT RFQ is provided by 20 pairs of water-cooled pi-rods. The pi-rods of 5 mm inner diameter and 10 mm of outer diameter, pass through 40 mm holes in the vanes[ 10 ]. To analysis the heat load on the pi-rods, a 520 mm pi-mode stabilizers period with 2 pairs of pi-rods is used to simulate. The power loss of each pi-rod is approximately 224 W. Fig. 6 shows the maximum temperature is 300.3 K, which indicates the pi-rods temperature is efficiently managed.

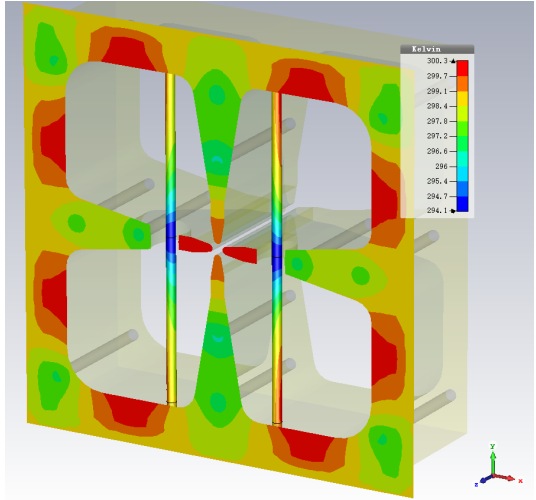

Fig. 6. Pi-rods temperature.

## 6 Tuners

The total 100 slug tuners are installed into the RFQ cavity. Each tuner has a diameter of 60 mm, with 20 mm nominal intrusion. The tuner period gives a tuning range of 1.4056 MHz, and the averaged power loss of a tuner is 73.8 W, which is small enough to go without water cooling. The heat load removal of each tuner depends on heat transfer to the RFQ cavity walls by thermal conduction. In the most critical case, the tuners with maximum penetration of 40 mm, shown in Fig. 7, are located in the extent of the cut-backs at RFQ entrance. The maximum tuner temperature change is 17.1 K, which is not problematical[11]. To evaluate the worst case, it provides a conservative estimation of the cooling capability to cool the tuner through RFQ body. It is likely that the other tuners will not experience such bad situation near the cut-backs. Thus, it is considered to be safe for operation.

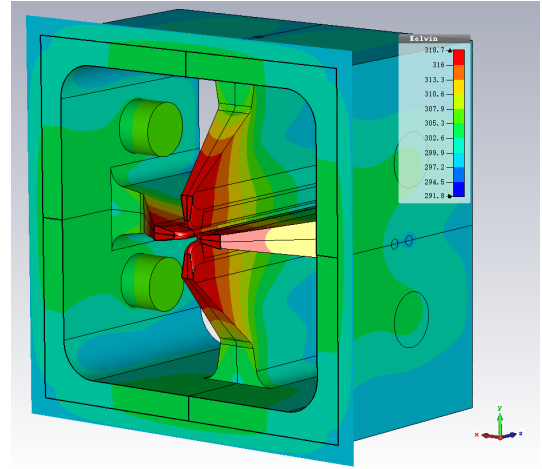

Fig. 7. Tuner temperature in the extent of the cut-backs at RFQ entrance.

## 7 Conclusion

A method of optimizing the layout of the cooling passages in the RFQ cross-section, evaluating the temperature field along the profile of vanes, has been proposed. The multiphysics results provide the prediction of RFQ operating parameters. The CST results and ANSYS results are in good agreement.